\documentstyle[aps,prl,twocolumn]{revtex}

\begin{document}

\draft
\wideabs{

\title{Environment-induced dynamical chaos}

\author{Bidhan Chandra Bag and Deb Shankar Ray}

\address{Indian Association for the Cultivation of Science,
Jadavpur, Calcutta 700 032, INDIA.}

\date{\today}

\maketitle

\begin{abstract}
We examine the interplay of nonlinearity of a dynamical system and 
thermal fluctuation of its environment in the ``physical limit'' of small 
damping and slow diffusion in a semiclassical context 
and show that
the trajectories of c-number variables exhibit dynamical chaos due 
to the thermal fluctuations of the bath.
\end{abstract} 

\pacs{PACS number(s): 05.45.Mt}
}

The interplay of nonlinearity of a dynamical system and thermal fluctuations 
of its environment has been one of the major areas of investigation in the
recent past.
These studies have enriched
our understanding of nonequilibrium processes in several contexts, such as,
symmetry between the growth and decay of classical fluctuations in equilibrium 
\cite{nature},
interesting topological features of patterns of paths of large fluctuations in nonlinear 
systems \cite{nature}, existence of generalized nonequilibrium potential  
\cite{tel}, influence of nonlinearity on dissipation in multiphoton 
processes \cite{gg2} and higher order diffusion in a nonlinear system 
\cite{batt}, etc.
While the development in these areas is largely confined to classical
domain we examine a related issue
in the semiclassical 
context. Since quantization is likely to add a new dimension to the interplay
of nonlinearity and stochasticity in a weakly dissipative system,
it is worthwhile to consider the physical limit of small
damping and slow diffusion due to thermal fluctuations of the environment
and look for the thermal 
fluctuations-induced features of nonlinearity in the dynamics.
In this communication we specifically explore some interesting aspects of 
dynamical chaos in a driven bistable system 
whose origin lies at the fluctuations of the environmental degrees of freedom.

To describe the dissipative quantum dynamics of a system we first consider the 
traditional system-reservoir model developed over the last few decades
\cite{gg2,louisell,gg1}. The
Hamiltonian of the bare system is coupled to an environment modeled by a 
reservoir of harmonic oscillator modes characterized by a frequency set 
$\{ \Omega_j \}$. The quantum dynamics is generated by  the overall Hamiltonian 
operator H for the system, environment and their coupling as follows;
\begin{equation}
H = H_0 + \hbar \sum_{j=1}^\infty \Omega_j {b_j }^\dagger b_j +
\hbar \sum_{j=1}^\infty 
\left[g(\Omega_j) b_j +g^\star (\Omega_j) {b_j}^\dagger \right] x ,
\end{equation}

\noindent
where
\begin{equation}
H_0 = \frac{p^2}{2 } + V(x) \; \;,
\end{equation}

\noindent
defines the usual  kinetic and  potential energy terms corresponding 
to the system, $x$ and $p$ being the position and momentum operators, respectively. 
The second and the third terms in (1) specify the reservoir modes and their linear
coupling to the system. $g (\Omega)$ denotes the system-reservoir coupling
constant.

Systematic elimination of the reservoir modes in the usual way, using Born and 
Markov approximations leads one to the following standard reduced 
density matrix  equation of the system only \cite{louisell}.
\begin{eqnarray}
\frac{d \rho}{dt} & = & - \frac{i}{\hbar} [H_0, \rho] 
+ \frac{\gamma}{2} (1 + \bar{n})
(2 a \rho a^\dagger - a^\dagger a \rho - \rho a^\dagger a) \nonumber \\
& & +\frac{\gamma}{2} \bar{n} 
(2a^\dagger \rho a - a a^\dagger \rho - \rho a a^\dagger) \; \;.
\end{eqnarray}

Here the system operator co-ordinate $x$ is related to the creation and 
the annihilation operators $a^\dagger, a$ respectively as 
$x = (\frac {1}
{\sqrt{2 \omega}}) (a + a^\dagger)$. $\omega$ is the linearised frequency
of the system.
Also the spectral density function of the reservoir is replaced by a continuous
density ${\cal D}(\omega)$. $\gamma > 0$ is the limit of  
$2 \pi |g(\omega)|^2 \frac{{\cal D}(\omega)}{\omega}$ as $\omega \rightarrow 0_+$ and is assumed
to be finite. $\gamma$ is the relaxation or dissipation rate and 
$\bar{n} \gamma$
is the diffusion coefficient $D$.
$\bar{n}(= [exp \left( \frac{\hbar \omega}{k T} \right)-1]^{-1})$
is the average thermal photon number of the reservoir. 
The terms analogous to Stark and Lamb shifts have been neglected. 

The first term in Eq.(3) corresponding to the
dynamical motion of the system refers to
Liouville flow. 
The terms containing $\gamma$ 
arise due to the interaction of the system with
the surroundings. 
While $\frac{\gamma \bar{n}}{2}$ terms denote the diffusion of fluctuations of
the reservoir modes into the system mode, $\frac{\gamma}{2}$ terms refer to the loss
of energy from the system into reservoir. In the limit $T\rightarrow 0, i. e. ,
\bar{n}\rightarrow 0$ the system is influenced by pure quantum noise or vacuum
fluctuations [ Note that 1 in $(\bar{n} + 1)$ in Eq.(3) corresponds to the
vacuum].

Our next task is to go over from a full quantum operator problem 
described by the Eq.(3) to an equivalent 
c-number problem . To this end 
we consider the 
quasi-classical distribution function W ($x, p$, t)
of Wigner. $x, p$ are now c-number variables. 
Rewriting Eq.(3) in a quasi-classical language we obtain
\begin{eqnarray}
\frac{\partial W}{\partial t} & = & 
-p \frac{\partial W}{\partial x}
+ \frac{\partial V}{\partial x} \frac{\partial W}{\partial p} 
+\frac{\gamma}{2} \eta \left(\frac{\partial x W}{\partial x} 
+ \frac{\partial p W }{\partial p}\right) \nonumber \\
& & +  \frac{\gamma \eta \hbar}{2 \omega}(\bar{n}+1)  
\frac{\partial^2 W}{\partial x^2} 
+ \frac{\gamma \eta \hbar \omega}{2} (\bar{n}+1)
\frac{\partial^2 W}{\partial p^2} \nonumber\\
& & + \sum_{n\geq 1}
\frac{ \hbar^{2n}(-1)^n}{ 2^{2n}(2n+1)! }
\frac{\partial^{2n+ 1}V}{\partial x^{2n+ 1} }
\frac{ \partial^{2n+1} W }{\partial p^{2n+1} }  \; \; .
\end{eqnarray} 

\noindent
$\eta$ in Eq.7 is a parameter used to identify the environment-induced
effect on the dynamics described by Eq.(4) (kept for bookkeeping in the 
calculation and put $\eta = 1$ at the end).

In the semiclassical limit $\hbar \omega << k T$, we have 
$\bar{n}+1 \approx \bar{n}$ and $D \approx \gamma k T$ 
so that Eq.(7) reduces to
\begin{eqnarray}
\frac{\partial W}{\partial t} & = & 
-p \frac{\partial W}{\partial x}
+ \frac{\partial V}{\partial x} \frac{\partial W}{\partial p} 
+\frac{\gamma}{2} \eta \left(\frac{\partial x W}{\partial x} 
+ \frac{\partial p W }{\partial p}\right) \nonumber \\
& & + D \eta \left(\frac{1}{\omega^2} \frac{\partial^2 W}{\partial x^2} 
+\frac{1}{2} \frac{\partial^2 W}{\partial p^2} \right ) \nonumber \\
& & + \sum_{n\geq 1}
\frac{ \hbar^{2n}(-1)^n}{ 2^{2n}(2n+1)! }
\frac{\partial^{2n+ 1}V}{\partial x^{2n+ 1} }
\frac{ \partial^{2n+1} W }{\partial p^{2n+1} }  \; \; .
\end{eqnarray} 

The overall dynamics described above is a superposition of two 
contributions, i. e. , the Liouville-Wigner dynamics and the system-reservoir 
dissipative dynamics. That the two contributions act independently is an 
assumption. 
The master equation (3) [or its Wigner function version (4)]
is the most popular one 
in quantum optics. It has been extensively used \cite{serg}, for the strongly 
nonlinear processes 
like three-wave, four-wave mixing and strong coherent light-matter interaction phenomena.
The equation has also been applied in the context of chaos, e. g., 
in the dissipative standard map \cite{dit}, dissipative logistic map \cite{goggin},
semiclassical theory of quantum noise in open chaotic systems \cite{bag1,bag2}
and in the studies of decoherence in relation to chaos for analysis of
quantum-classical correspondence \cite{zurek,akp}.

In the semiclassical ($\hbar \rightarrow 0$)
limit the dissipative quantum dynamics can
be conveniently described by ``WKB-like'' ansatz (we refer to `` WKB-like''
since we are considering more that one dimension. Traditional WKB refer to 
one dimension only)
\cite{nature,bo} of Eq.(5) for Wigner function of the form
\begin{equation}
W(x, p, t) = Z(t) \exp(-\frac{s}{\hbar}) \; \;.
\end{equation}

Here $Z(t)$ is a prefactor and $s(x, p, t)$ is a classical action which is a
function of c-number variables $x$ and $p$ , satisfying the following Hamilton-Jacobi equation
\begin{eqnarray}
& & \frac{\partial s}{\partial t} + p \frac{\partial s}{\partial x} - 
\frac{\partial s}{\partial x} \frac{\partial s}
{\partial p} - \frac{\gamma}{2} \eta (x \frac{\partial s}{\partial x} + 
p \frac{\partial s}{\partial p}) 
+ \eta (\frac{\partial s}{\partial x})^2 \nonumber\\
& & + \eta \omega^2 (\frac{\partial s}{\partial p})^2 +
\sum_{n\geq 1} \frac{x^{2n} (-1)^{3n+1}}{2^{2n}(2n)!} \frac{\partial^{2n+1} V}
{\partial x^{2n+1}} \frac{\partial s}{\partial p} = 0 \; \;.
\end{eqnarray}

The derivation of Eq.(7) is based on the consideration of the ``physical limit'' 
of weak dissipation and slow fluctuations in the sense
$\frac{D_1}{\hbar^2} \approx \frac{1}{\hbar}$ where $D_1=\frac{D}{2 \omega^2}$
(note that $D_1$ and $\hbar$ have same dimension).
The above equation can be solved by integrating the Hamiltonian equations of
motion,

\begin{eqnarray}
\dot{x} & = & p - \frac{\gamma}{2} \eta x + 2 \eta P 
\nonumber\\ 
\dot{X} & = & P -  \frac{\gamma}{2} \eta X 
\nonumber\\
\dot{p} & = & V' + \frac{\gamma}{2} \eta p - 2 \omega^2 \eta X -
\sum_{n\geq 1}  \frac{ (-1)^{3n+1}}{2^{2n}(2n)!} 
\frac{\partial^{2n+1} V}{\partial x^{2n+1}}  X^{2n} \nonumber\\
\dot{P} & = & V'' X + \frac{\gamma}{2} \eta P - 
\sum_{n\geq 1} \frac{(-1)^{3n+1}}{2^{2n}(2n+1)!} 
\frac{\partial^{2n+2} V}{\partial x^{2n+2}} X^{2n+1}
\end{eqnarray}

\noindent
which are derived from the following Hamiltonian $H_{eff}$
\begin{eqnarray}
H_{eff} & = & p P - V' X 
-\frac{\gamma}{2} \eta (x P+p X) +\eta (P^2 +\omega^2 X^2)   \nonumber\\
& & + 
\sum_{n\geq 1} \frac{(-1)^{3n+1}}{2^{2n}(2n+1)!} 
\frac{\partial^{2n+1} V}{\partial x^{2n+1}}  X^{2n+1} \; \; .
\end{eqnarray} 

\noindent
Here we have put $\frac{\partial s}{\partial x} = P$ and
$\frac{\partial s}{\partial p} = X$.
The introduction of additional degree-of-freedom by incorporating the 
auxiliary momentum (P) the and co-ordinate (X) makes the system effectively a 
two-degree-of-freedom system. The origin of these two variables is the thermal
fluctuations 
of the reservoir. It is easy to identify the environment-related terms containing
$\eta$ in Eqs.(7-9). The auxiliary Hamiltonian (9) is therefore not to be 
confused with the microscopic Hamiltonian (1) which describes a system in 
contact with a reservoir with infinite degrees of freedom.
Although the phase space trajectories concern fluctuations of c-number variables
in the formal sense, because of the equations of motion (8) described
by a Hamiltonian (9), the motion is strictly deterministic.
The experiments on the corresponding classical version of the problem by
Luchinsky and McClintock \cite{nature} have demonstrated that a trajectory of fluctuation is 
indeed a part of
physical reality. We  emphasize, however, here a number of distinguishing
features in this context. While the studies by Luchinsky and McClintock
\cite{nature} and
Graham and Tel \cite{tel} concern overdamped limit, we  consider here a weakly dissipative
system. Furthermore because of the quantum correction, 
the phase space trajectories
of fluctuations are significantly modified by semiclassical features.
The introduction
of these {\it quantum features at a semiclassical level} through a c-number Hamiltonian description of a
dissipative evolution in the physical limit of weak damping and slow
diffusion due to thermal noise
is the essential content of the present work. 
Since the Eq.(9) describes deterministic evolution under nonlinear potential, the
pattern of trajectories of fluctuations may display chaotic behaviour. In what
follows we investigate this dynamical aspect 
of the dissipative system.

The equations derived in the weak thermal noise limit for the weakly dissipative 
semiclassical systems are fairly general. For illustration we now consider a simple 
model system Hamiltonian $H_0$ (see Eq.2)
\begin{equation}
H_0 = \frac{p^2}{2} + a x^4 - b x^2 + g x cos \Omega_0 t
\end{equation}

which describes a bistable potential driven by a time-periodic field. $a$ and $b$
are the constants of the potential $V(x)$. The fourth term
in (10) includes the effect of coupling of the system as well as the strength of 
the field of frequency $\Omega_0$.
For the Hamiltonian (10)  the Eqs.(8) read as
\begin{eqnarray}
\dot{x} & = & p - \frac{\gamma}{2} \eta x + 2 \eta P
\nonumber\\ 
\dot{X} & = & P -  \frac{\gamma}{2} \eta X 
\nonumber\\
\dot{p} & = & 4 a x^3 -2 b x +g \cos{\Omega}t  
+ \eta (\frac{\gamma}{2} p - 2 \omega^2 X) -3 a x X^2 \nonumber\\
\dot{P} & = & (12 a x^2 -2 b) X + \frac{\gamma}{2} \eta P - a X^3
\end{eqnarray}

\noindent
which are derivable from the auxiliary Hamiltonian
\begin{eqnarray}
H_{eff} & = & Pp -( 4 a x^3 -2 b x + g \cos{\omega} t) X
-\frac{\gamma}{2} \eta (x P + p X ) \nonumber \\
& & + \eta (P^2 + \omega^2 X^2) + a x X^3  \; \; .
\end{eqnarray}

The system Hamiltonian (10) has served as a standard paradigm for a number of
theoretical and experimental investigations \cite{nature,bag1,lin}
over many years. For the present purpose we choose the following parameter
values; $a = \frac{1}{4}$, $b = \frac{1}{2}$, $\Omega_0 = 6.07$ and $g = 10$.
Since we are considering the physical limit of weak damping  and small
diffusion we take the initial
conditions for the auxiliary 
variables X and P (which originate from the fluctuations due to environment) as
$P = 0$ and $X\rightarrow 0$ (we have used $X = 1.5 \times 10^{-6}$). 
This ensures a vanishing Hamiltonian $H_{eff}$
for the entire numerical investigation that follows below.

We first consider a specific trajectory with the initial condition $p = 0$
and $x = -2.512$ for small values of $\gamma$ (typical $0.01$).
Under this condition($\eta = 1$) the system is vanishingly coupled to the
surroundings and consequently the dynamical behaviour is 
effectively due to the weak dissipation only.
We illustrate this situation in Fig. 1 in terms of
a Poincare map for the phase space which exhibits 
strong global chaos. On the other hand when the parameter $\eta$ is switched off 
($\eta = 0$) the system displays typical weak chaos (Fig.2). 
The similar behaviour has been observed for other sets of initial
conditions for $x$ and $p$ (We have not reproduced them here for the sake of
brevity).

The effect of weak dissipation and slow diffusion due to thermal 
fluctuations from the surroundings can be seen in the case of other sets of 
initial conditions also. For the initial condition $p = 0$ and $x = -2.49$
one observes for $\gamma = 0.01$ dissipative strong chaos($\eta = 1$).
This is illustrated in Fig.3. It is interesting to note that for $\eta = 0$
the same trajectory gets localized in the left well as a regular one as shown 
in Fig.4. The weakly chaotic and regular trajectories in Figs. 2 and 4, respectively,
are purely 
semiclassical in nature (in the absence of any coupling to the surroundings).
The strong global chaotic behaviour as shown in Figs. 1 and 3 has therefore its
origin in the thermal fluctuations of the reservoir.
In other words the chaotic behaviour or its enhancement is exclusively due to thermal
fluctuations from the surroundings, which becomes appreciable even in the 
physical limit of weak damping and slow diffusion of the
thermal noise within the semiclassical description. We have checked this assertion for 
some other values of the
initial conditions for the system oscillator.

The reduction of the system-reservoir Hamiltonian description [H in Eq.1]
for a dissipative quantum system to an auxiliary Hamiltonian 
description [$H_{eff}$ in Eq.9] 
effectively reduces the infinite-degree-of-freedom
system to a two-degree-of-freedom system where the auxiliary
degree of freedom characterized by X and P owe their origin in the fluctuations
of the reservoir. Since X and P appear as the multiplicative factors in the 
auxiliary Hamiltonian $H_{eff}$, the weak thermal noise limit makes $H_{eff}$ 
a vanishing Hamiltonian. 
The observed semiclassical chaos
may therefore be regarded as a dynamical manifestation of the interplay of
nonlinearity and thermal fluctuations.

In this paper we have examined the weak thermal noise limit of a semiclassical 
dissipative nonlinear system. We have shown that the vanishing Hamiltonian
method 
can be suitably extended to follow
the phase space trajectories of fluctuations of c-number variables which
exhibit dynamical chaos.
In view of the accessibility 
of the model to analogue electronic circuits \cite{nature} we believe that the results
discussed bear further experimental relevance 
in the semiclassical 
context. 

\noindent
B. C. Bag is indebted to the CSIR for partial financial support and to 
J. Ray Chaudhuri for helpful discussions.

\begin{figure}
\caption{Plot of x vs p on the Poincare surface of section ($X= 0$) for Eq.(11) with initial
condition $x = -2.512 $, $p = 0$, $X \rightarrow 0$, $P = 0$, $\gamma = 0.01$
$\eta = 1$. (Units are arbitrary).}
\end{figure}

\begin{figure}
\caption{Same as in Fig.1 but for $\eta = 0$.}
\end{figure}

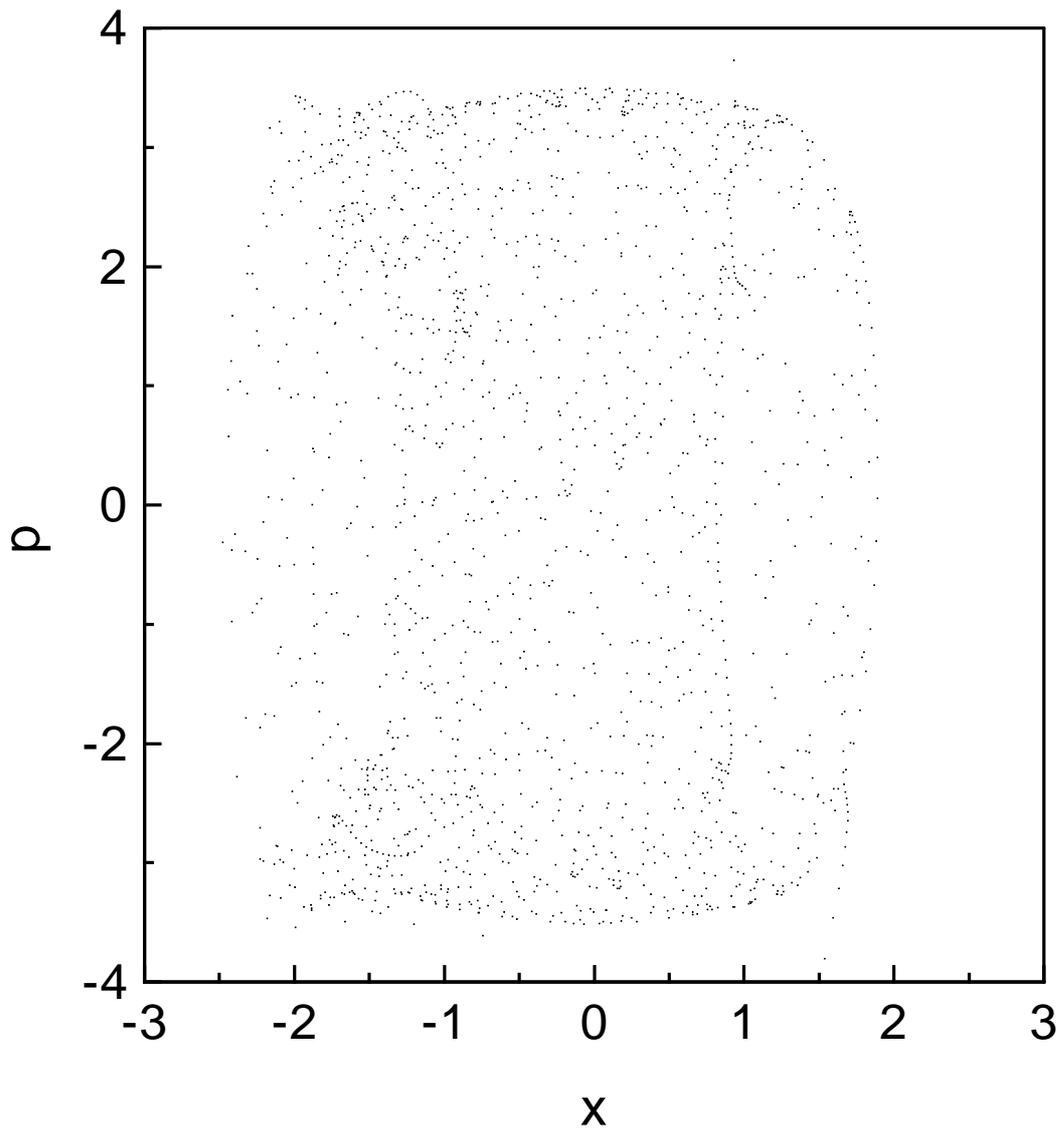
\begin{figure}
\caption{Same as in Fig.1 but for $x =-2.49$.}
\end{figure}

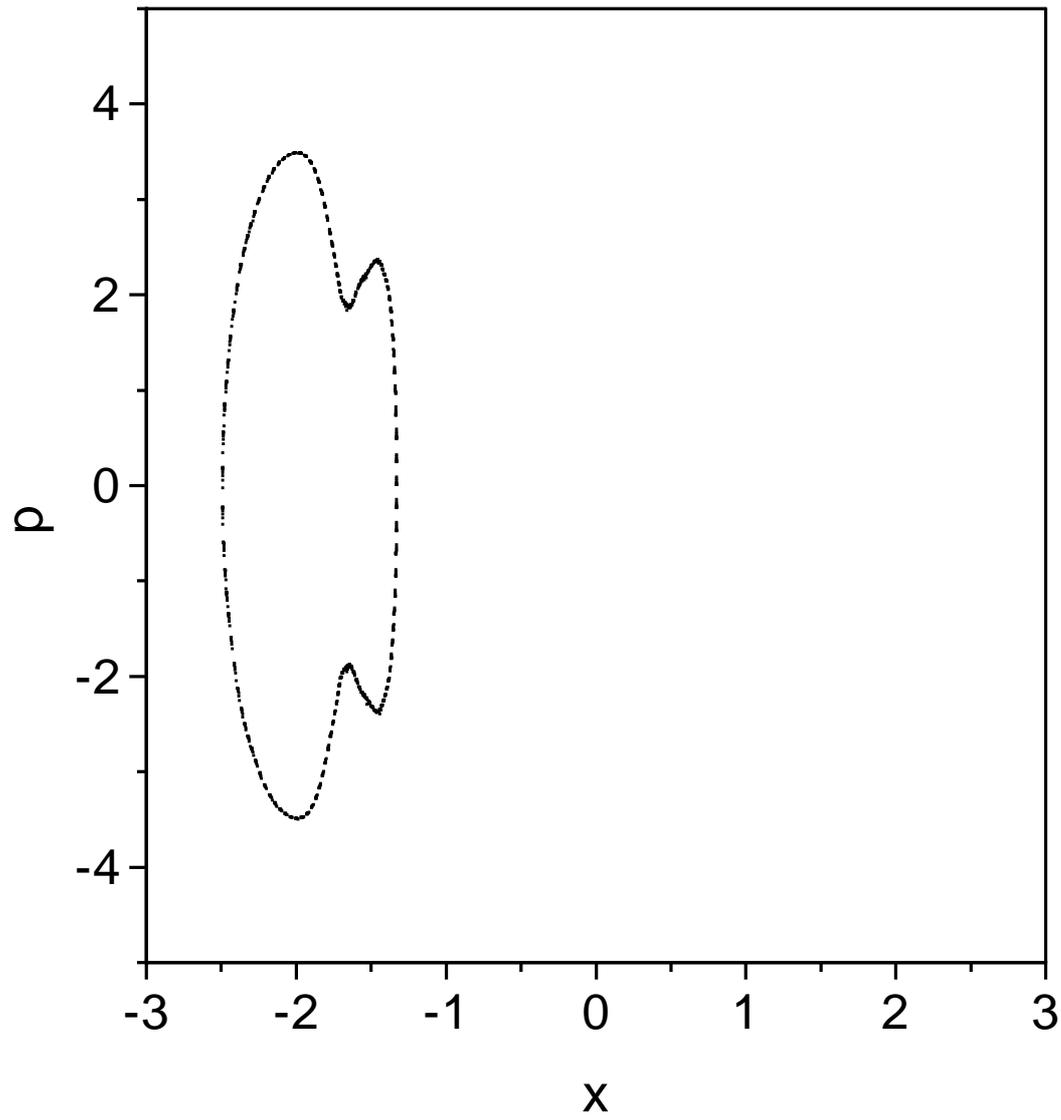
\begin{figure}
\caption{Same as in Fig.3 but for $\eta=0$.}
\end{figure}

\end{document}